# NONLINEAR BEHAVIOUR OF TIME-STEPPING ALGORITHMS FOR INITIAL VALUE PROBLEMS

Stefano Sello[1]

## ABSTRACT

Recent advances in nonlinear dynamical systems theory provide a new insight into numerical properties of discrete algorithms developed to solve nonlinear initial value problems. Basic features like accuracy and stability are well pointed out through diagrams or maps of computed asymptotic solutions in a suitable parametric space. Applying this methodology to a nonlinear test equation, we compared some numerical features of the well known second-order Crank-Nicolson solver with those of a recent proposed version which is fourth-order accurate. The approach gives some useful indication on the capabilities of familiar and innovative ODE integrators when applied to nonlinear problems.

[1]CISE- Tecnologie Innovative, Applied Mathematics Section
Via Reggio Emilia 39 - 20090 Segrate MILANO (Italy)
Tel. 39+2+21672216 Fax 39+2+21672620 E-mail 284allse@alliant.cise.it

1. INTRODUCTION

It is well consolidated the strong interaction existing between dynamical systems and numerical analysis. Numerical properties of time-stepping algorithms require an extension to the classical analysis performed in the linear domain, especially when we deal with applications to a large class of nonlinear problems arising in physics and engineering. Despite of a consolidated literature concerning the theory of nonlinear properties of numerical methods (see, e.g. [1] to a nonlinear stability analysis for Runge-Kutta solvers) often we need more practical tools of analysis to a global evaluation of the reliability and capability of selected discrete numerical schemes for the time integration of nonlinear systems of ordinary differential equations (ODE).

The recent advances in the techniques of nonlinear dynamical systems, offer a new insight into the numerical behaviour of discrete dynamical systems, i.e. the result of a given combination of the initial continuous differential problem and the discrete numerical scheme employed to resolve it. The interest and efficiency of such approach is well documented in literature (see, e.g. [2] for an extended paper on the subject), and the related area of applied mathematics is fast growing ("dynamics of numerics").

The purpose of the present note is to apply the techniques of analysis typical of the dynamics of numerics to compare the (nonlinear) numerical performances of the classical second-order Crank-Nicolson algorithm with a recently proposed version of it which achieves the fourth-order accuracy, while retaining (in the limit case of a linear and Hamiltonian case) the stability, conservative properties and structure implementation of the underlying second-order method [3].

## 2. METHODOLOGY OF ANALYSIS

The essence of the dynamics of numerics approach is the concept that the dynamics of discrete systems, arising from the combination of continuous models and selected numerical algorithms, is only loosely related to that of continuous systems. Previous analyses show, indeed, that the dynamics of discrete systems is far richer and complex than their continuous counterparts, and the involved computations can create spurious solutions (not necessarily divergent) [2].

The analysis starts with a continuous model, or test problem, with known explicit solutions, followed by a discretization of it according to a chosen (standard or not) numerical algorithm. Following [2] here we consider as a continuous problem an ordinary differential test equation, namely the *logistic equation* [4]:

$$\begin{cases} \dfrac{dy(t)}{dt} = \alpha y(t)[1-y(t)] = f(y(t)) \\ y(t_0) = y_0 \end{cases} \qquad (1)$$

where $\alpha$ is a given parameter (here >0). Further, we apply the techniques from discrete dynamics to analyze the numerical behaviour of the classical unconditionally stable second-order Crank-Nicolson solver, coming from the "trapezoidal rule",[5], (Nonautonomous case):

$$y_{n+1} = y_n + \frac{\Delta t}{2}[f(y_n;t_n) + f(y_{n+1};t_{n+1})] \qquad (2)$$

where $\Delta t$ is the integration time step and $t_n = t_0 + n\Delta t$; and a related fourth-order accurate version proposed in [3] as a three successive fractional steps procedure:

$$\begin{aligned} 1)\ y_{n+\beta} &= y_n + \frac{\beta \Delta t}{2}[f(y_n;t_n) + f(y_{n+\beta};t_{n+\beta})] \\ &= y_n + \beta \Delta t\ \tilde{f}(y_n,y_{n+\beta};t_n,t_{n+\beta}) \end{aligned}$$

$$2)\ y_{n+1-\beta} = y_{n+\beta} + (1-2\beta)\Delta t\ \tilde{f}(y_{n+\beta}, y_{n+1-\beta}; t_{n+\beta}, t_{n+1-\beta})$$

$$3)\ y_{n+1} = y_{n+1-\beta} + \beta\Delta t\ \tilde{f}(y_{n+1-\beta}, y_{n+1}; t_{n+1-\beta}, t_{n+1}) \tag{3}$$

where β is a "universal" constant given by: $\beta=1/3(2+2^{(-1/3)}+2^{(1/3)})$. Note that the second fractional step is backward in time. Of course, both the methods are implicit in nature and, thus, require an iterative convergence procedure for the numerical computation.

When applied to the logistic equation (1) the solvers (2) and (3) produce asymptotic solutions that coincide (or not) with the exact fixed points of the discretized map. In particular, the quadratic Bernouilli equation (1) leads to the following analytical solution:

$$y(t) = \frac{y_0}{y_0 + (1-y_0)e^{-\alpha(t-t_0)}} \tag{4}$$

where $y_0$ denotes the initial condition. Analytically it is trivial to determine the two fixed points, $y^*$, for the solution (4): $y_1^*=1$ (linearly stable); $y_2^*=0$ (linearly unstable).

To explore the "whole" dynamical behaviour of the above solvers, when applied to discretized version of eqn.(1), we peformed an extensive parametric computation of the asymptotic solutions varying either the initial condition, $y_0$, and the value of the "control" parameter: $r=\alpha\Delta t$. (Note that here $\alpha$ is the "Jacobian" of the linearized equation.) Through the computation of the related maps in the parametric space (bifurcation diagrams) it is possible to display the "global" behaviour of the numerical schemes in terms of its collection of asymptotic solutions [4]. In this way we can point out the capability of the schemes (when applied to nonlinear problems) quantified by the presence (or not) of spurious solutions (divergent or not) of the discretized model. In many cases the spurious steady states lie beyond the linearized stability limit (if exists), but in some cases the numerical algorithm can produce spurious solutions also *below* this limit generating evident trouble. (See [2] for an account of these features in many standard algorithms.)

Finally, note from eqn.(1) that the system can be considered "dissipative" when y>1/2 because in this case the "trace of the Jacobian" is lesser than zero, being the conservative condition verified in the point: y=1/2.

## 3. NUMERICAL COMPUTATIONS

The eqn.(1) was numerically integrated using the second-order Crank-Nicolson solver, (2), and its extended fourth-order accurate version, (3), through an extensive parametric exploration of their asymptotic solutions. As example, Figure 1 shows the result for the bifurcation diagram of a "pseudo-logistic" map, i.e. the map arising from the application to eqn.(1) of the (conditionally stable) explicit Euler solver [4]:

$$y_{n+1} = y_n + \Delta t\, f_n(y_n; t_n) = y_n[1 + \Delta t\, \alpha(1 - y_n)] \tag{5}$$

As we see, the stable fixed point, $y^* = 1$, is the unique numerical solution found by the scheme (5) *if we stay below* the linearized stability limit at r=2. Beyond that point a bifurcation occurs and the map starts to exhibit stable spurious solutions of period higher than one (multiple solutions at a given control parameter r). If the r value is sufficiently high (arising from an increase of $\alpha$, $\Delta t$ or both) a nonperiodic type solution occurs covering quite uniformly the available solution space ("chaotic regime"). Of course, beyond the stability limit there is also the possibility to find divergent numerical solutions, with a catastrophic impact on the computation. Note in Figure 1 the absence of plotted points beyond the value r=3, due to the presence of divergent solutions only.

In this work the bifurcation diagrams have been produced dividing the control r axis into 200 equal intervals and, for each r value, we explored a set of different initial conditions (typically 5 or 10). To avoid an initial transient we discarded the first 800 time steps and successively we recorded and plotted the remaining 200 iterations assuming, as in [2], that there are not practical differences between a stable fixed point of period 200 and a nonperiodic solution. The above means that each bifurcation diagram displays about 200000 solution points. Even if these diagrams are a course approximation of the true bifurcation picture, nevertheless they involve a quite expensive computational effort, especially in the case of implicit numerical schemes.

For a comparison, Figure 2 shows the equivalent bifurcation diagram with the use of a very efficient A-stable semi-explicit Runge-Kutta type solver, which is third-order accurate ("RRCM-asymptot", [6]). Note, in this case, the absence (for the r range explored) of spurious (stable or unstable) solutions. This result, as we will see, is not trivial because we handle a nonlinear problem.

Figure 3 shows the computed bifurcation diagram for the classical Crank-Nicolson solver, and in figure 4 the equivalent diagram for the fourth-order accurate version. The numerical computation for the second-order method required about 11 hours of CPU-time on a IBM RISC/6000 work-station (file size ~ 8 Mb); whereas the more expensive fourth-order version required about 40 hours of CPU-time.

As we can see from figures, the bifurcation diagrams look very different: for the second-order method, if we stay below the linearized stability limit of the explicit scheme (5) (r=2), the only numerical solution computed is the "exact" stable fixed point; whereas the fourth-order version displays spurious solutions even well below the stability limit. More precisely, we detected a degradation of the computed solution just beyond the value r=1.16 and, more important, we found also divergent solutions, depending on the selected initial condition ($y_0$>1), for each value of the control r. (There is an exception inside the window: 1.4<r<1.495, where no unstable solutions are present.)

The first bifurcation with a cascade of spurious (stable) solutions occurs at r=2 for the second-order method; whereas the splitting to a multiple solutions with periods greather than one occurs at r=1.51 for the fourth-order version, also with regions (below r=2) displaying a non periodic type behaviour (i.e. covering quite uniformly an interval of the parametric space).

In the interval 2<r<3.2 the classical second-order method displays stable spurious solutions only; on the other hand beyond the last value, there is an onset of divergent solutions, depending on the initial condition. For the fourth-order version the picture is more critical because we found that, beyond the linear stability limit, r=2, the only solutions detected are unstable, i.e. generated by divergent numerical computations. Indeed no stable (spurious or not) solutions have been detected, beyond the value r=2. (See in figure 4 the absence of plotted points when r>2.) Finally we note that below the limit value r=2 the unstable, divergent fixed points detected was 26.3% of the total computed solutions.

A plausible explanation of the critical behaviour of the fourth-order version is the presence, in its procedure, of a fractional step backward in time that produces severe troubles, especially when dealing with (nonlinear) non conservative problems [3]. The computed bifurcation diagrams point out this feature in an impressive way. The above analysis allows a better evaluation of the numerical performances of algorithms selected to solve initial value problems, especially in nonlinear domains.

## 4. CONCLUDING REMARKS

The dynamical behaviour of numerical methods, gives us a valuable tool in order to point out the nonlinear features of numerical integrators designed to solve initial value problems. Previous extensive analyses on the subject show that bifurcations to and from spurious (divergent or not) solutions can appear beyond and also below linearized stability limits of the selected numerical schemes. In this work the above approach of analysis was applied to compare the nonlinear features of the classical second-order Crank-Nicolson solver and a related extension which is fourth-order accurate. The computed bifurcation diagrams allow a better evaluation of the accuracy and stability properties, and their mutual relations, when we deal with equations in a nonlinear domain.

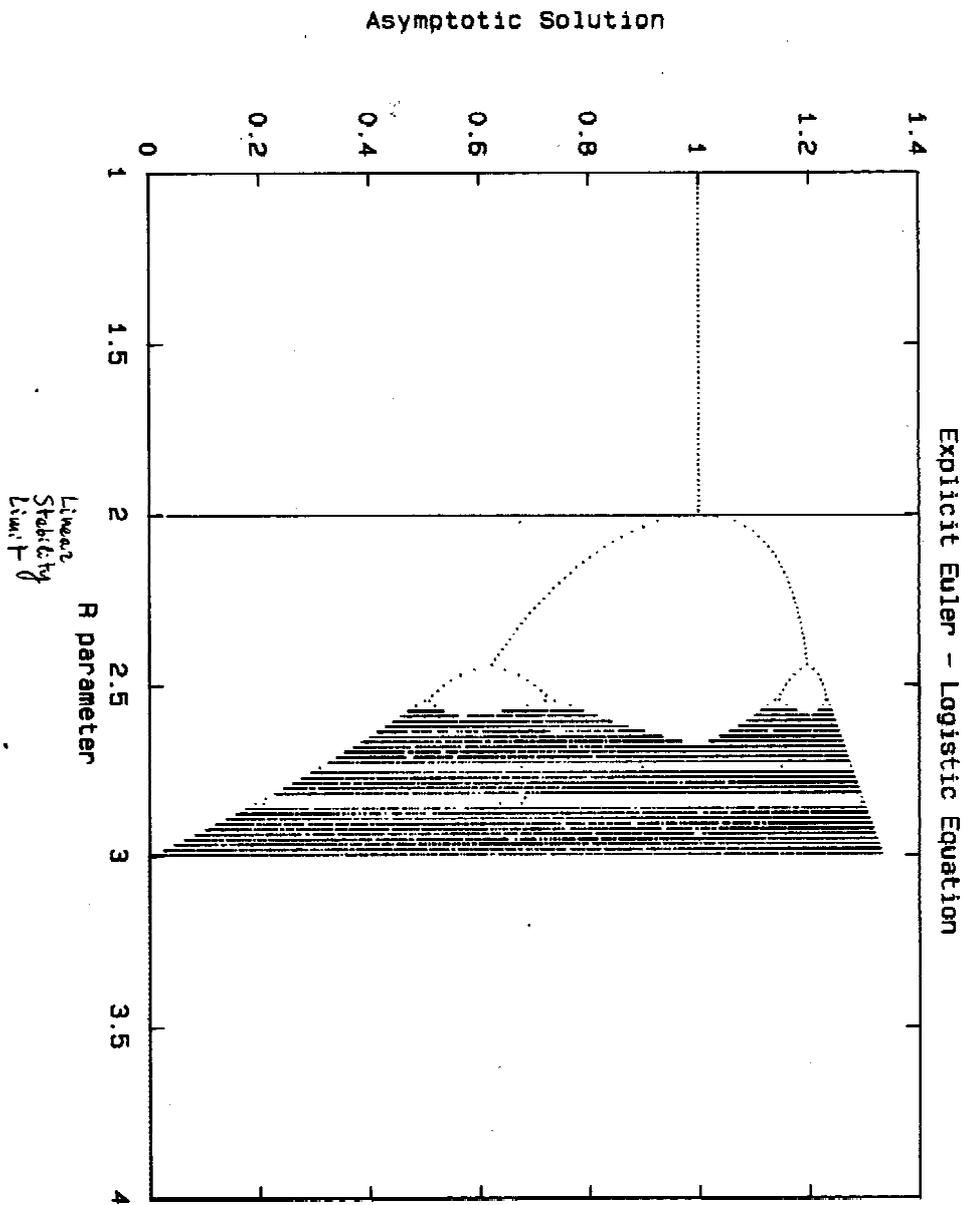

Figure 1

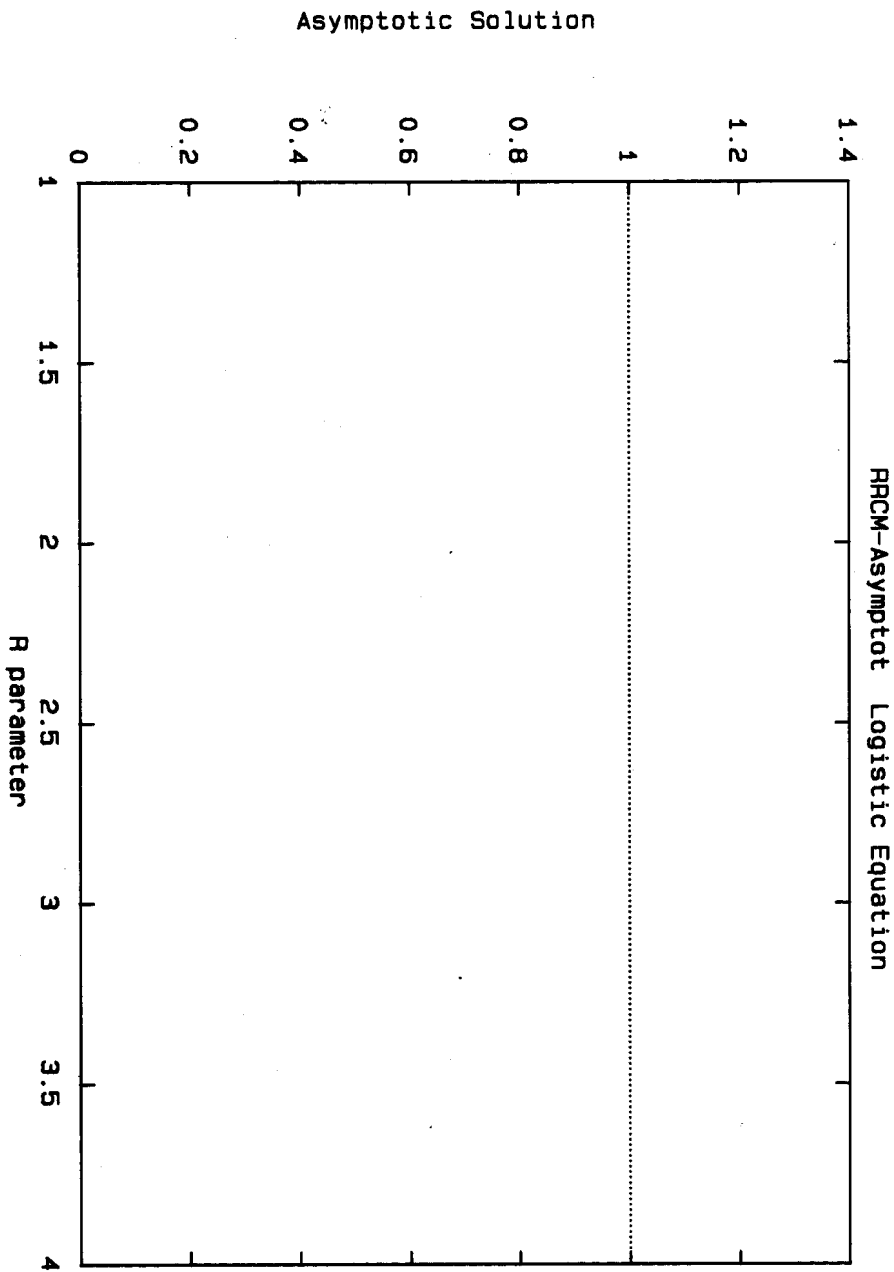

Figure 2

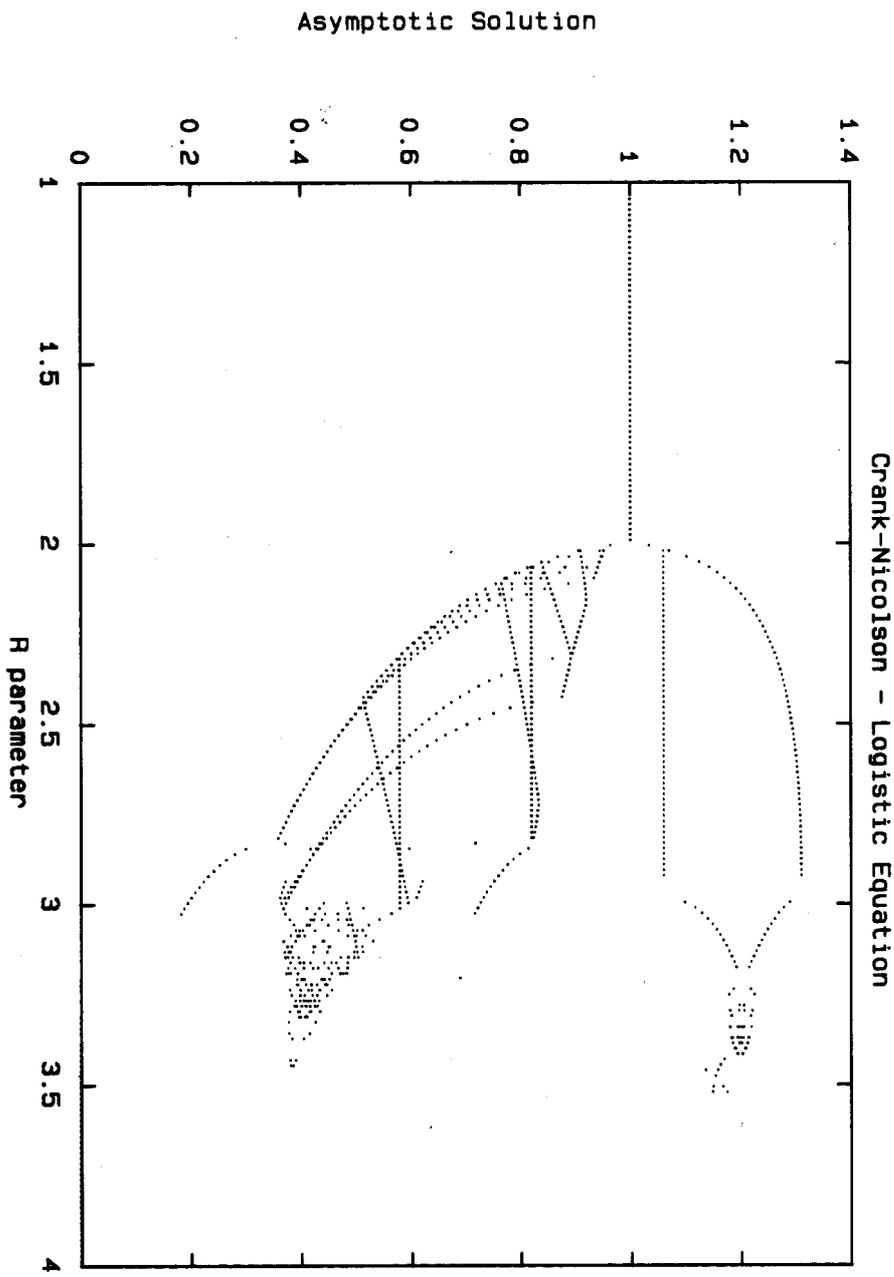

Figure 3

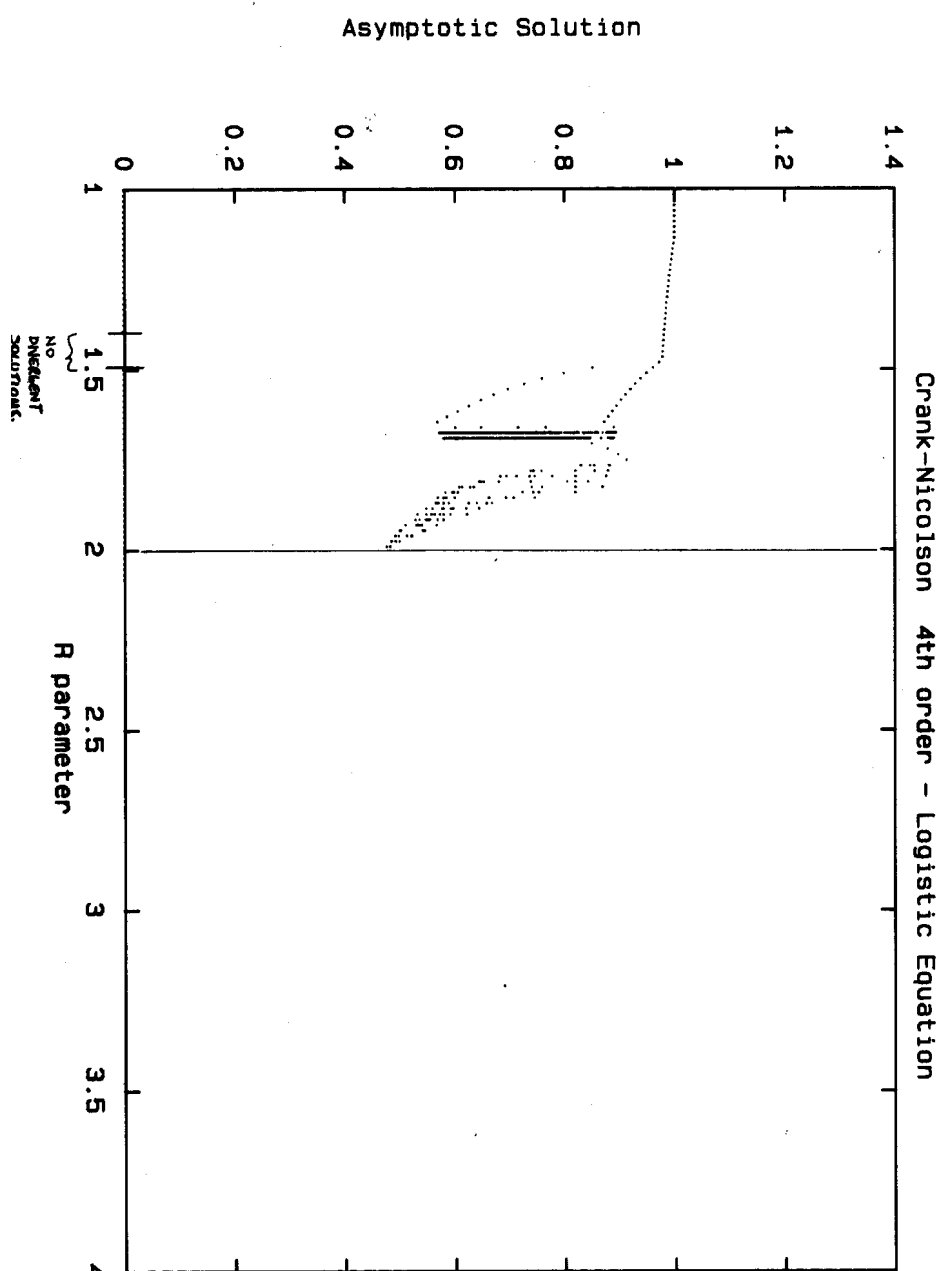

Figure 4